\documentclass[cameraready]{Interspeech}

\usepackage{pgfplots}
\usepackage{pgfplotstable}
\pgfplotsset{compat=1.18}
\title{Probing Low Frame Rate Degradation in Neural Audio Codecs}

\author[affiliation={1}, orcid=0000-0002-2226-9882]{Alex}{Gichamba}
\author[affiliation={1}, orcid=0000-0002-5245-0113]{Moise}{Busogi}

\address{
    $^1$ Carnegie Mellon University Africa, Rwanda
}

\email{angicham@andrew.cmu.edu, mbusogi@andrew.cmu.edu}


\usepackage{comment}

\begin{document}

\maketitle

\begin{abstract}
Low frame rates in neural audio codecs are attractive for autoregressive speech synthesis, where the generation cost scales linearly with the sequence length. Recent work has demonstrated that codecs can operate at 12.5 Hz and below, but the mechanisms underlying low frame rate degradation remain insufficiently understood. We investigate these mechanisms through a controlled frame rate ablation. We reproduce a quality cliff at 6.25 Hz reported in previous works and evaluate candidate explanations: phonemic collisions and codebook saturation, neither of which shows evidence of a fundamental barrier. The cliff is instead caused by suboptimal training configuration: fixed clip duration during training yields too few tokens at low frame rates, starving the decoder of inter-token context. Once corrected, WER degrades smoothly with phonemic load down to 3.1 Hz and 1.6 Hz, suggesting the inference-time efficiency gains of low frame rate codecs are more accessible than previously assumed.
\end{abstract}

\section{Introduction}

Neural audio codecs \cite{zeghidour_soundstream_2021} compress continuous audio waveforms into sequences of discrete tokens. Beyond their applications in audio compression, modern speech processing pipelines rely on neural audio codecs as speech tokenizers, serving as a bridge
between raw audio and the language modeling frameworks that underpin
text-to-speech synthesis systems \cite{wang_neural_2023,
hu_qwen3-tts_2026} and spoken dialogue systems
\cite{defossez_moshi_2024}.

The modern neural audio codec paradigm can be largely attributed to
SoundStream \cite{zeghidour_soundstream_2021} and EnCodec
\cite{defossez_high_2023}, both pairing a fully convolutional
encoder-decoder with Residual Vector Quantization (RVQ) and training
end-to-end with adversarial, feature matching, and multi-scale spectral
reconstruction losses. DAC \cite{kumar_high-fidelity_2023} addressed
codebook collapse through factorized codes and L2-normalized
quantization, and introduced Snake activations and a multi-band STFT
discriminator to push fidelity to state-of-the-art levels. Subsequent
work considerably diversified the design space: SpeechTokenizer
\cite{zhang_speechtokenizer_2023} distilled HuBERT semantics into the
first RVQ layer to disentangle content from paralinguistic information;
Mimi \cite{defossez_moshi_2024} adopted a split-RVQ design with
transformer bottleneck layers to achieve streaming-compatible 12.5 Hz
tokenization; SNAC \cite{siuzdak_snac_2024} introduced multi-scale RVQ
operating at multiple temporal resolutions simultaneously; and BigCodec
\cite{xin_bigcodec_2024} abandoned RVQ in favour of a single
8192-entry codebook, demonstrating that scaling model capacity beyond
150M parameters can compensate for the representational constraints of
single-level VQ at ultra-low bitrates.

Practitioners specifying a codec for speech synthesis face several key
design decisions. The frame rate $f_r$, the duration represented by
each token, has a direct and significant impact on downstream
generation cost. Recent speech synthesis systems such as
Moshi~\cite{defossez_moshi_2024} and Qwen3-TTS~\cite{hu_qwen3-tts_2026}
adopt a two-stage autoregressive pipeline: a primary decoder
autoregressively generates the token sequence for the 0th codebook,
after which a secondary decoder fills the remaining residual levels
conditioned on it. The cost of the primary decoder scales linearly with
sequence length and therefore linearly with frame rate. At 12.5 Hz a
10-second utterance requires 125 autoregressive steps; at 50 Hz the
same utterance requires 500 steps. Reducing frame rate is therefore a
direct lever on inference latency and throughput, independent of model
size, architecture, or quantization depth.

Despite this incentive, very low frame rates have been associated with
significant quality degradation, and the mechanisms behind this
degradation have remained unclear. \cite{casanova_nanocodec_2025}
trained codecs at 25, 12.5, and 6.25 Hz and observed that while the
25 Hz and 12.5 Hz models performed comparably, the 6.25 Hz model showed
a significant decline particularly in intelligibility. They hypothesized that
this was because at 6.25 Hz the model frequently compresses two distinct
phonemes within a single frame , consistent with English speakers
producing approximately 10--12 phonemes per second , and concluded that
phonemic collision sets a practical lower bound on codec frame rates.
Subsequent work addressed low frame rate quality through increasingly
complex architectural solutions: FlexiCodec \cite{li_flexicodec_2025}
introduced dynamic frame rate allocation with ASR feature-assisted
dual-stream encoding and Transformer bottlenecks to preserve semantic
information at rates as low as 3 Hz.

\begin{figure}[ht]
\centering
\pgfplotsset{
    std/.style={
        color=blue!70!black,
        mark=*,
        mark size=1.5pt,
        line width=1pt,
    },
    matched/.style={
        color=orange!60!black,
        mark=square*,
        mark size=1.5pt,
        line width=1pt,
        dashed,
    },
    phonemes/.style={
        color=gray!70!black,
        mark=triangle*,
        mark size=1.5pt,
        line width=1pt,
        dotted,
    },
}

\begin{tikzpicture}
\begin{axis}[
    width=7.5cm, height=5.5cm,
    xmode=log,
    log ticks with fixed point,
    xtick={1.6, 3.125, 6.25, 12.5, 25, 50, 100},
    xticklabels={,,,,,,,,},
    xmin=1.2, xmax=130,
    y tick label style={font=\footnotesize},
    ylabel={WER (\%) $\downarrow$},
    ylabel style={font=\small},
    ymin=0, ymax=115,
    ytick={0,25,50,75,100},
    grid=both,
    grid style={dotted, gray!40},
    axis lines=left,
    tick align=outside,
    legend pos=north east,
    legend style={font=\footnotesize, draw=none, fill=white, fill opacity=0.8},
]
\addplot[std] coordinates {
    (6.25,  107.4)
    (12.5,  10.62)
    (25,    5.90)
    (50,    5.38)
    (100,   5.10)
};
\addplot[matched] coordinates {
    (1.6,   63.22)
    (3.125, 29.36)
    (6.25,  15.37)
    (25,    5.79)
    (50,    5.38)
    (100,   5.02)
};
\legend{Fixed $T_\text{clip}$,Fixed $K$}
\end{axis}
\end{tikzpicture}

\vspace{-0.3cm}

\begin{tikzpicture}
\begin{axis}[
    width=7.5cm, height=4.0cm,
    xmode=log,
    log ticks with fixed point,
    xtick={1.6, 3.125, 6.25, 12.5, 25, 50, 100},
    xticklabels={1.6, 3.1, 6.25, 12.5, 25, 50, 100},
    xmin=1.2, xmax=130,
    x tick label style={font=\footnotesize, rotate=45, anchor=east},
    y tick label style={font=\footnotesize},
    xlabel={$f_r$ (Hz)},
    xlabel style={font=\small},
    ylabel={phones/frame},
    ylabel style={font=\small},
    ymin=0.9, ymax=6.5,
    ytick={1.0, 2.0, 3.0, 4.0, 5.0, 6.0},
    grid=both,
    grid style={dotted, gray!40},
    axis lines=left,
    tick align=outside,
    legend pos=north east,
    legend style={font=\footnotesize, draw=none, fill=white, fill opacity=0.8},
]
\addplot[phonemes] coordinates {
    (1.6,   6.197)
    (3.125, 3.987)
    (6.25,  2.595)
    (12.5,  1.794)
    (25,    1.357)
    (50,    1.128)
    (100,   1.010)
};
\legend{phones/frame}
\end{axis}
\end{tikzpicture}

\caption{Top: WER of DAC variants under fixed clip duration (solid
blue) and fixed token sequence length (dashed orange). The quality
cliff at 6.25 Hz under fixed clip duration disappears when sequence
length is matched across frame rates, after which WER degrades
smoothly and monotonically with phonemic load (bottom). Bottom: mean
phonemes per frame from MFA forced alignments on LibriSpeech
\texttt{test-clean}, extending to 3.125 Hz and 1.6 Hz where
intelligible speech persists at bitrates as low as 192 bps.}
\label{fig:wer_phones}
\end{figure}
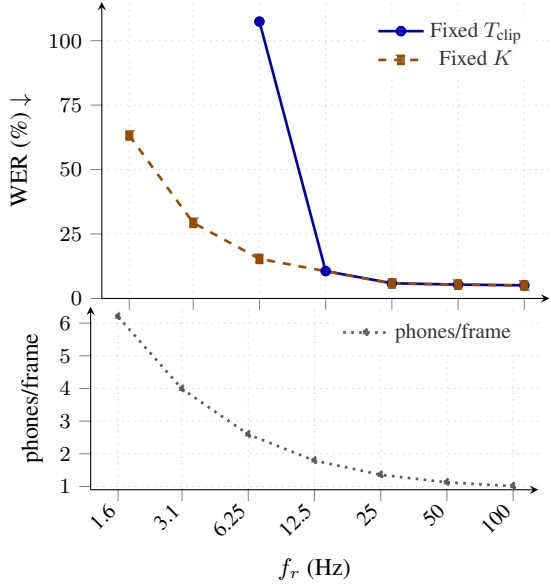

In this work we ask a more fundamental question: \textit{why} does
standard codec training fail at low frame rates, and is the failure
intrinsic to the frame rate itself? Through a controlled ablation of
the DAC framework~\cite{kumar_high-fidelity_2023} spanning 1.6 to
100 Hz, we identify a training misconfiguration as the primary cause:
standard training maintains a fixed clip duration regardless of frame
rate, yielding too few tokens per example at low frame rates and
preventing the decoder from learning inter-token coherence. We evaluate
phonemic collision and codebook saturation as alternative explanations
and find no evidence of a fundamental barrier in either case
(Figure~\ref{fig:wer_phones}). Once the misconfiguration is corrected,
reconstruction intelligibility degrades smoothly with phonemic load,
suggesting that residual degradation at very low frame rates reflects expected information loss due to the limited representational capacity.
This extends to 3.125 Hz and 1.6 Hz without architectural
modifications, confirming that the inference-time efficiency gains of
low frame rate codecs are more accessible than previously assumed.

\section{Preliminaries}
\label{sec:preliminaries}
\subsection{General RVQ Codec Formulation}
We borrow from the notation in \cite{wu_towards_2025} in our codec modeling 
below. Consider an input waveform of single channel audio to be encoded as a 
one dimensional signal $\boldsymbol{w} \in \mathbb{R}^{Tf_s}$ sampled at a 
frequency $f_s$ Hertz and of duration $T$ seconds.

Neural audio codecs encode input waveforms into compressed discrete 
representations through a two-stage process. We first define the encoder as a 
function $E: \mathbb{R}^{Tf_s} \rightarrow \mathbb{R}^{Tf_r \times d}$ that 
maps the input waveform to a continuous hidden representation, where $f_r$ is the frame rate in Hertz (typically much smaller 
than $f_s$), $Tf_r$ is the number of frames in the compressed representation, and $d$ is the dimensionality of the hidden representation.

The hidden representation $\boldsymbol{h}$ is then processed by a Residual 
Vector Quantizer (RVQ) $Q: \mathbb{R}^{Tf_r \times d} \rightarrow 
\mathcal{V}^{Tf_r \times n_q}$ that discretizes the representations in the latent space.\ $\mathcal{V} := \{1, 2, \ldots, |\mathcal{V}|\}$ is the discrete 
vocabulary of size $|\mathcal{V}|$, and $n_q$ is the number of codebook levels. 
The RVQ operates residually letting each successive level quantize the residual of the previous level so that level 0 captures the coarsest representation and each subsequent level refines the residual error.

Correspondingly, we define the decoder as a function $D: \mathcal{V}^{Tf_r 
\times n_q} \rightarrow \mathbb{R}^{Tf_s}$ that approximately reconstructs the encoded waveform.

The theoretical bit rate is the minimum information rate required to transmit 
the quantized representation. Since we have $f_r$ frames per second, each 
containing $n_q$ quantized codes from a vocabulary of size $|\mathcal{V}|$, 
the bit rate is:
\begin{equation}
\label{eq:bit_rate}
    R = f_r \cdot n_q \cdot \log_2 |\mathcal{V}| \quad \text{bits per second}
\end{equation}

\subsection{Codebook Utilization and Entropy Efficiency}
Let $\mathbf{Z}^{(q)} = \{z_{t,q}\}$ denote the multiset of all level-$q$ code 
indices emitted by the encoder across the evaluation corpus. The \textit{codebook 
utilization} at level $q$ is the fraction of vocabulary entries assigned to at least one frame, expressed as:
\begin{equation}
    U_q = \frac{\left|\left\{ z_{t,q} \right\}\right|}{|\mathcal{V}|}
\end{equation}
Let 
$\hat{p}_q(v) = |\{t : z_{t,q} = v\}| \,/\, |\mathbf{Z}^{(q)}|$ be the 
empirical probability of code $v$ at level $q$. The \textit{entropy efficiency} 
is:
\begin{equation}
    \eta_q = \frac{H(\hat{p}_q)}{\log_2 |\mathcal{V}|}
    \quad \text{where} \quad
    H(\hat{p}_q) = -\sum_{v \in \mathcal{V}} \hat{p}_q(v) \log_2 \hat{p}_q(v)
\end{equation}
A value of $\eta_q = 1$ indicates uniform usage of the full codebook, 
representing maximum information capacity per token. A value approaching $0$ 
indicates codebook collapse, where the encoder maps the majority of inputs to 
a small set of codes.

\section{Experimental Setup}

\subsection{Reference Codecs}
To situate our ablation within the broader landscape of neural audio 
codecs, we evaluate a set of published pretrained models without any 
retraining. For 16 kHz speech we include DAC-16k \cite{kumar_high-fidelity_2023} and BigCodec~\cite{xin_bigcodec_2024}, which replaces RVQ 
with a single large codebook at 80 Hz. For 24 kHz speech we evaluate 
DAC-24k, Mimi~\cite{defossez_moshi_2024}, SNAC~\cite{siuzdak_snac_2024}, 
WavTokenizer~\cite{ji_wavtokenizer_2025}, and the Qwen3-TTS 12.5 Hz speech 
tokenizer~\cite{hu_qwen3-tts_2026}. All models are evaluated on LibriSpeech 
\texttt{test-clean} \cite{panayotov_librispeech_2015}. Model outputs are resampled to the target frequency for each metric that operates at a specific frequency.

\subsection{Evaluation Metrics}
Reconstructed audio is assessed across four dimensions. 
\subsubsection{Intelligibility}
\textbf{Word Error Rate (WER)}: is computed by transcribing 
reconstructed audio with MMS-1B~\cite{pratap_scaling_2023} and measuring word error rate against reference transcriptions, serving as a model-based measure of intelligibility.\\
\textbf{Short-Time Objective Intelligibility (STOI)}~\cite{taal_short-time_2010}: predicts speech intelligibility by computing the average correlation between clean and processed speech across short-time time-frequency regions.
\subsubsection{Voice Preservation}
\textbf{Speaker Similarity (SPK-SIM)}: We compute the cosine similarity between the speaker embeddings of the reference and decoded audios. We use a  WAVLM-based \cite{chen_wavlm_2022} model finetuned for speaker recognition via Espnet-SPK \cite{jung24c_interspeech}.
\subsubsection{Distortion}
\textbf{Mel Cepstral Distortion (MCD)} \cite{kubichek_mel-cepstral_1993}: measures spectral distance between reference and synthesized speech by computing the Euclidean distance between mel-frequency cepstral coefficients (MFCCs).

\subsubsection{Pseudo Mean Opinion Scores}
\textbf{UTMOS} \cite{saeki_utmos_2022}:  is a learned mean opinion 
score predictor that estimates perceptual naturalness without a reference signal, scored 1--5.

\subsection{Frame Rate Study}
All models in our controlled frame rate study are based on the 16 kHz Descript Audio Codec 
(DAC)~\cite{kumar_high-fidelity_2023}. The encoder consists of $L$ strided convolutional blocks, 
each containing three dilated residual units, followed by a Snake activation~\cite{ziyin_neural_2020}. The decoder mirrors this structure with transposed convolutions.

The frame rate $f_r$ is determined by the total stride product of the 
encoder, which must satisfy $\prod_{l=1}^{L} s_l = f_s / f_r$ for $f_s = 
16{,}000$ Hz. We construct codec variants spanning a wide range 
of frame rates by varying the per-block strides while keeping all other 
architectural parameters fixed. The decoder strides are the reverse of the encoder  strides in all cases. Bitrate is computed from Equation~\ref{eq:bit_rate} with $n_q = 12$ and $|\mathcal{V}| = 1024$ held constant, so that $R = 120 f_r$ bps and bitrate varies solely as a function of frame rate.

\subsection{Training}
All ablation models are trained on LibriSpeech \texttt{train-clean-100}~\cite{panayotov_librispeech_2015}, a read English speech dataset sampled at 16 kHz. All models are trained for 100{,}000 iterations on a single NVIDIA H100-80 GPU using the Adam optimizer with the learning rate schedule from the original DAC implementation.

Following the DAC training method, clips are drawn by randomly cropping each utterance to a fixed duration $T_{\text{clip}}$ seconds, yielding $K = \lfloor T_{\text{clip}} 
\cdot f_r \rfloor$ tokens per training example. In the baseline 
configuration we follow DAC and set $T_{\text{clip}} = 0.38$ seconds for 
all variants. This results in substantially fewer tokens per clip at lower frame rates: as few as $K = 2$ at 6.25 Hz compared to $K = 19$ at the 50 Hz baseline. To isolate the effect of this disparity, we additionally retrain models with a fixed $K$.

\section{Results}

\subsection{Performance of Reference Codecs}

Table~\ref{tab:reference} reports performance metrics for published and
publicly available neural audio codecs evaluated on LibriSpeech
\texttt{test-clean}. Among 16 kHz models,
DAC-16k at its default 50 Hz configuration achieves a WER of 5.25\% at
6{,}000~bps, serving as a strong upper-bound reference. BigCodec, which
replaces RVQ with a single 8192-entry codebook, achieves a WER of 7.77\%
at a substantially lower bitrate of 1{,}040~bps, demonstrating that
single-codebook designs can remain competitive at low bitrates.

Among 24 kHz models, DAC-24k at 75 Hz achieves near-perfect STOI of 0.99
and WER of 5.01\% at 24{,}000~bps. Mimi, which was specifically engineered
for low frame rate tokenization via a transformer bottleneck and split-RVQ
design, achieves STOI of 0.96 and WER of 5.52\% at just 1{,}100~bps and
12.5 Hz, the strongest result among low-bitrate models. WavTokenizer-75
achieves comparable STOI of 0.90 at a similar bitrate of 900~bps, though
with higher WER of 11.37\%. Models operating below 50 Hz without
architectural accommodations for low frame rates show clear degradation:
SNAC at its multi-scale rates and WavTokenizer-40 at 40 Hz both exhibit
substantially higher WER (13.14\% and 24.36\% respectively) and reduced
speaker similarity, consistent with the quality cliff we investigate in
the following sections. Notably, Qwen3-TTS-Tokenizer achieves competitive
WER of 5.43\% and high SPK-SIM of 0.91 despite an anomalously low STOI
of 0.65, reflecting its design objective of preserving semantic content
and speaker identity over waveform fidelity.

\begin{table*}[t]
\small
\setlength{\tabcolsep}{4pt}
\centering
\caption{Reconstruction quality of published neural audio codecs on
LibriSpeech \texttt{test-clean}. $f_r$ denotes frame rate in Hz,
$n_q$ is the number of quantization levels, $|\mathcal{V}|$ is
codebook size, and $R$ is bitrate. Bold denotes best within each
frequency group.}
\label{tab:reference}
\begin{tabular}{lccccccccc}
\toprule
Model & $f_r$ (Hz) & $n_q$ & $|\mathcal{V}|$ & $R$ (bps)
& STOI $\uparrow$ & WER (\%) $\downarrow$
& MCD $\downarrow$ & SPK-SIM $\uparrow$ & UTMOS $\uparrow$ \\
\midrule
\multicolumn{10}{l}{\textit{16 kHz}} \\
DAC-16k~\cite{kumar_high-fidelity_2023}
    & 50 & 12 & 1024 & 6000
    & \textbf{0.97} & \textbf{5.25} & \textbf{2.04} & \textbf{0.90} & 4.00 \\
BigCodec~\cite{xin_bigcodec_2024}
    & 80 &  1 & 8192 & 1040
    & 0.93 & 7.77 & 3.01 & 0.81 & \textbf{4.11} \\
\midrule
\multicolumn{10}{l}{\textit{24 kHz}} \\
DAC-24k~\cite{kumar_high-fidelity_2023}
    & 75 & 32 & 1024 & 24000
    & \textbf{0.99} & \textbf{5.01} & \textbf{1.02} & \textbf{0.94} & 4.06 \\
Qwen3-TTS-Tok.~\cite{hu_qwen3-tts_2026}
    & 12.5 & 16 & 2048 & 2200
    & 0.65 & 5.43 & 7.11 & 0.91 & \textbf{4.13} \\
Mimi~\cite{defossez_moshi_2024}
    & 12.5 &  8 & 2048 & 1100
    & 0.96 & 5.52 & 2.49 & 0.88 & 3.92 \\
SNAC~\cite{siuzdak_snac_2024}
    & 12,23,47 & 3 & 4096 & 984
    & 0.88 & 13.14 & 4.07 & 0.59 & 3.02 \\
WavTokenizer-75~\cite{ji_wavtokenizer_2025}
    & 75 &  1 & 4096 & 900
    & 0.90 & 11.37 & 3.92 & 0.66 & 3.79 \\
WavTokenizer-40~\cite{ji_wavtokenizer_2025}
    & 40 &  1 & 4096 & 480
    & 0.85 & 24.36 & 4.99 & 0.52 & 3.57 \\
\bottomrule
\end{tabular}
\end{table*}

\subsection{Frame Rate Ablation}
\label{sec:ablation}

Table~\ref{tab:frame_rate_results} reports reconstruction quality across
all frame rate variants trained with the standard clip duration
$T_{\text{clip}} = 0.38$s. Performance degrades gradually as frame rate
decreases from 100 Hz to 12.5 Hz, with modest but consistent reductions
across all metrics. At 12.5 Hz the codec remains competitive, achieving
STOI of 0.89 and WER of 10.62\% at 1{,}500~bps. However, reducing the
frame rate by a further factor of two to 6.25 Hz produces a catastrophic
collapse: STOI drops to 0.46, WER increases to 107.4\%, MCD rises to
20.17, and SPK-SIM falls to 0.09. This sharp discontinuity, a
$10\times$ increase in WER for a $2\times$ reduction in frame rate,
motivates the analysis in the following subsections.

\begin{table}[htbp]
\footnotesize
\setlength{\tabcolsep}{4pt}
\centering
\caption{Performance metrics across frame rate variants trained
with standard clip duration $T_{\text{clip}} = 0.38$s.
$\dagger$ denotes the DAC baseline trained with the same parameters as variants for a controlled comparison.}
\label{tab:frame_rate_results}
\begin{tabular}{lccccc}
\toprule
$f_r$ (Hz) & STOI $\uparrow$ & WER (\%) $\downarrow$
& MCD $\downarrow$ & SPK-SIM $\uparrow$ & UTMOS $\uparrow$ \\
\midrule
100   & 0.98 &   5.10 &  1.33 & 0.97 & 4.03 \\
50$^\dagger$
      & 0.97 &   5.38 &  2.00 & 0.93 & 3.98 \\
25    & 0.95 &   5.90 &  2.61 & 0.86 & 3.88 \\
12.5  & 0.89 &  10.62 &  4.06 & 0.62 & 3.02 \\
6.25  & 0.46 & 107.40 & 20.17 & 0.09 & 1.27 \\
\bottomrule
\end{tabular}
\end{table}

\subsection{Phoneme Collisions}
\label{sec:collision}

A natural first hypothesis is that the quality cliff reflects a
fundamental theoretic limit. At 6.25 Hz each frame spans
160~ms of audio, and given an average phoneme rate of 12.2 phonemes
per second in the test set, each frame must encode, on average, 2.6
distinct phonemes, a potential cause hypothesized by \cite{casanova_nanocodec_2025} as the primary cause of intelligibility collapse. The bottom panel of Figure~\ref{fig:wer_phones} confirms that phonemes per frame increases
monotonically with decreasing frame rate, consistent with this
hypothesis.

However, the top panel reveals a strong decoupling between phonemic
load and WER. Under fixed token sequence length, $K$ training, the 6.25 Hz model encodes the same 2.6 phonemes per frame yet achieves WER of 15.37\% rather than 107.4\%, and intelligible speech persists down to 1.6 Hz where each frame spans over six phonemes on average. Phonemic collision is therefore a correlate of the performance cliff rather than its cause.

\subsection{Codebook Utilization}
\label{sec:codebook}

A second hypothesis is that codebook saturation explains the cliff: at
low frame rates the encoder must map diverse 160~ms acoustic contexts
into a fixed vocabulary of 1024 codes, potentially exhausting the
codebook's representational capacity. We test this by computing
codebook utilization $U_q$ and entropy efficiency $\eta_q$ at each
frame rate as defined in Section~\ref{sec:preliminaries}.

Table~\ref{tab:codebook} shows that both metrics are essentially flat
across all frame rates. Utilization remains above 98.7\% and entropy
efficiency varies by less than 0.05 across the full range of
frame rates, including at 6.25 Hz. The encoder produces well-distributed
codes regardless of frame rate, and the codebook operates near capacity
in all conditions. Codebook saturation is therefore ruled out as an
explanation for the performance cliff.

\begin{table}[htbp]
\footnotesize
\setlength{\tabcolsep}{4pt}
\centering
\caption{Codebook utilization $U_0$ and entropy efficiency $\eta_0$ at
RVQ level 0 across frame rate variants. At levels 1--11, all models
achieve $\eta_q > 0.90$ with differences across frame rates of less
than 0.01.}
\label{tab:codebook}
\begin{tabular}{lcc}
\toprule
$f_r$ (Hz) & $U_0$ & $\eta_0$ \\
\midrule
6.25  & 0.987 & 0.822 \\
12.5  & 0.997 & 0.817 \\
25    & 0.999 & 0.837 \\
50    & 1.000 & 0.866 \\
100   & 1.000 & 0.868 \\
\bottomrule
\end{tabular}
\end{table}

\subsection{Training Sequence Lengths as the Root Cause}
\label{sec:fix}

Having ruled out phoneme collision and codebook saturation as
fundamental barriers, we turn to the training procedure. Under the
standard configuration, all models are trained on clips of
$T_{\text{clip}} = 0.38$ seconds, following the DAC baseline. At 50 Hz
this yields $K = 19$ tokens per training clip, sufficient for the
decoder to observe multiple token boundaries and learn inter-token
coherence. At 6.25 Hz the same clip duration yields only $K = 2$
tokens per clip, meaning the decoder is trained almost exclusively on
single-token reconstruction and never learns to produce coherent audio
across token boundaries. At inference on full-length utterances, the
decoder must produce sequences of 40--50 tokens, a regime it has never
encountered during training.

To test this hypothesis, we retrain the models with a fixed sequence length ($K=19$), matching the 50 Hz
baseline. Comparing the 6.25 Hz row in Table~\ref{tab:frame_rate_results} against the corresponding row in
Table~\ref{tab:main_results}, this single change recovers performance
substantially: STOI improves from 0.46 to 0.89, WER drops from
107.40\% to 15.37\%, MCD falls from 20.17 to 3.72, and SPK-SIM
recovers from 0.09 to 0.62, bringing the 6.25 Hz model to near parity
with the 12.5 Hz baseline at half the bitrate. The performance cliff is therefore not a consequence of any information-theoretic limit imposed by low frame rates.

\subsection{Extension to Ultra-Low Frame Rates}
\label{sec:extension}

The recovery of 6.25 Hz performance raises a natural question: how low
can the frame rate go while still producing intelligible speech,
provided the training clip duration is matched appropriately? We train
two additional models at 3.125 Hz and 1.6 Hz using the
matched-token protocol ($K = 19$ tokens per clip).

Table~\ref{tab:main_results} reports results for all matched-token
models including the ultra-low frame rate variants. At 3.125 Hz, where
each token spans 320~ms and the bitrate falls to just 375~bps, the
model achieves STOI of 0.84 and WER of 29.36\%, a substantial
degradation relative to 6.25 Hz but far from the catastrophic collapse
observed under standard training. At 1.6 Hz, where each token spans 625~ms and the bitrate falls to 192~bps, STOI remains at 0.76 and
WER at 63.22\%, indicating that the codec retains meaningful
intelligibility at compression ratios that would have been considered
entirely infeasible under the standard training protocol.

\begin{table}[htbp]
\footnotesize
\setlength{\tabcolsep}{4pt}
\centering
\caption{Reconstruction quality across frame rate variants trained
with matched sequence length $K = 19$ ($T_{\text{clip}} = 19 / f_r$).
$\dagger$ denotes the DAC baseline.}
\label{tab:main_results}
\begin{tabular}{lccccc}
\toprule
$f_r$ (Hz) & STOI $\uparrow$ & WER (\%) $\downarrow$
& MCD $\downarrow$ & SPK-SIM $\uparrow$ & UTMOS $\uparrow$ \\
\midrule
100   & 0.98 &  5.02 & 1.45 & 0.96 & 3.99 \\
50$^\dagger$
      & 0.97 &  5.38 & 2.00 & 0.93 & 3.98 \\
25    & 0.95 &  5.79 & 2.50 & 0.90 & 3.99 \\
12.5  & 0.93 &  7.17 & 3.06 & 0.82 & 4.00 \\
6.25  & 0.89 & 15.37 & 3.72 & 0.62 & 3.80 \\
3.125 & 0.84 & 29.36 & 4.63 & 0.48 & 3.24 \\
1.6   & 0.76 & 63.22 & 5.67 & 0.32 & 2.67 \\
\bottomrule
\end{tabular}
\end{table}

\section{Conclusion}
We investigated why neural audio codecs degrade at low frame rates,
asking whether the failure is intrinsic to the frame rate itself.
Phonemic collision and codebook saturation show no evidence of a
fundamental barrier. The degradation instead resulted from a training
misconfiguration. However, when sequence length is fixed during training,
reconstruction intelligibility degrades smoothly with phonemic load,
suggesting that residual degradation at very low frame rates reflects expected
information loss due to the limited representational capacity. This
extends to 3.125 Hz and 1.6 Hz at bitrates as low as 192 bps,
showing that the inference-time efficiency gains of low frame rate
codecs are more accessible than previously assumed.

\section{Acknowledgments}
This publication was developed as part of the African Engineering and Technology Network (Afretec), which is managed by Carnegie Mellon University Africa and receives financial support from the Mastercard Foundation.  The views expressed in this document are solely those of the authors and do not necessarily reflect those of Carnegie Mellon University Africa or the Mastercard Foundation.

This work used Bridges2 at Pittsburgh Supercomputing Center through allocation CIS250700 from the Advanced Cyberinfrastructure Coordination Ecosystem: Services \& Support (ACCESS) program \cite{10.1145/3569951.3597559}, which is supported by U.S. National Science Foundation grants \#2138259, \#2138286, \#2138307, \#2137603, and \#2138296.

\section{Generative AI Use Disclosure}
Generative AI tools were used in two limited capacities in this work. First, they were used to develop portions of the code that supported our experiments. All generated code was reviewed and validated by the authors. Second, they were used to identify grammatical errors in the manuscript and suggest corrections. All outputs were reviewed and approved by the authors.

\bibliographystyle{IEEEtran}
\bibliography{mybib}

\end{document}